\documentclass[pre,twocolumn,superscriptaddress,floatfix]{revtex4-2}
\usepackage{amsmath,graphicx,amssymb,url,color}
\usepackage[utf8]{inputenc}
\usepackage[T1]{fontenc}
\usepackage[english]{babel}

\newcommand{\la}{\left\langle}
\newcommand{\ra}{\right\rangle}

\begin{document}
\title{Solitary states in spiking oscillators with higher-order interactions}
\author{Vladimir V. \surname{Semenov}}\email{semenov.v.v.ssu@gmail.com}
    \affiliation{Department of Physics, Saratov State University, Astrakhanskaya str. 83, 410012 Saratov, Russia}
\author{Subhasanket \surname{Dutta}}
    \affiliation{Complex Systems Lab, Department of Physics, Indian Institute of Technology Indore, Indore 452020, India}
\author{Stefano \surname{Boccaletti}}
    \affiliation{Sino-Europe Complexity Science Center, North University of China, Taiyuan 030051, China}
    \affiliation{Institute of Interdisciplinary Intelligent Science, Ningbo University of Technology, Ningbo, China}
    \affiliation{CNR -- Institute of Complex Systems, Madonna del Piano 10, Sesto Fiorentino, Firenze 50019, Italy}
\author{Charo I. \surname{del Genio}}
    \affiliation{Institute of Interdisciplinary Intelligent Science, Ningbo University of Technology, Ningbo, China}
    \affiliation{Institute of Smart Agriculture for Safe and Functional Foods and Supplements, Trakia University, Stara Zagora 6000, Bulgaria}
    \affiliation{School of Mathematics, North University of China, 030051, Taiyuan, China}
\author{Sarika \surname{Jalan}}
    \affiliation{Complex Systems Lab, Department of Physics, Indian Institute of Technology Indore, Indore 452020, India}
\author{Anna \surname{Zakharova}}
    \affiliation{Bernstein Center for Computational Neuroscience, Humboldt-Universität zu Berlin, Philippstraße 13, 10115 Berlin, Germany}

\begin{abstract}
We study a system of globally coupled FitzHugh-Nagumo
oscillators with pairwise and second-order interactions,
showing that the presence of higher-order interactions substantially
affects the character of the transition between synchronous
and asynchronous states driven by changes in coupling strength.
In particular, we demonstrate that, around the synchronization
transition, solitary states emerge due to the presence of
second-order interactions. We also show that,
in difference to the phenomenology observed in systems of phase oscillators,
at low fixed pairwise coupling strengths, solitary states appear for both
transition directions as the second-order interaction strength is varied,
whereas for higher couplings they only occur in the
forward direction, with the backwards one characterized by
explosive desynchronization.
\end{abstract}

\maketitle

\section{Introduction}
In the past decade, the presence of higher-order interactions
has been recognized as having a substantial impact on the behaviour
of a large number of complex systems~\cite{Bat20,Bat22}. Their
effects are particularly relevant because in the general case
they cannot be recovered via the superposition of pairwise interactions.
This has revived interest in the study of the
static and dynamical properties of the structures that represent
them, namely hypergraphs and simplicial complexes~\cite{Boc23,Ma25,Par25,del25}.

The influence of higher-order interactions on collective dynamics
has been investigated in particular depth for networks of oscillators,
where their relation to phenomena such as synchronization and resonance
has been well established~\cite{Tla19,Gal22,Par22,Sem25}. In such
systems, one of their most striking effects is that they can give
rise to abrupt \emph{desynchronization} transitions in the absence
of abrupt \emph{synchronization} counterparts. Additionally, they
can induce extreme multistability, whereby infinitely many stable
and partially synchronized states coexist~\cite{Ska19}. Such coexistence
relates to the remarkable phenomenon of solitary states, in which
one or more dynamical units split off and behave differently from
the others despite the coupling being homogeneous~\cite{Jar18,Maj19,Hel20,Sch21}.

Classically, solitary states appear at the edge of synchrony along the
desynchronization path, and can be precursors for the emergence of
chimera states~\cite{Jar15,Sch22}. Their characteristic dynamical features
are largely determined by the coupling properties.
For example, in the presence of adaptive coupling, the transition to desynchronization
accompanied by the appearance of solitary states can occur via the
emergence of multiclusters with hierarchical structure in cluster size~\cite{Ber20,Ber21}.
Also, in globally coupled networks of excitable systems, where repulsion dominates over attractive interactions,
solitary states in non-locally coupled arrays inherit their dynamical properties from unbalanced
periodic two-cluster states~\cite{Fra22}.

Notably, the formation of chimera states can also be an effect
of the partial synchronization caused by higher-order interactions~\cite{Kun22},
which are additionally known to induce
abrupt synchronization transitions with hysteresis
and bistability of synchronized and incoherent states~\cite{Ska20,Kac22},
ultimately causing explosive synchronization~\cite{Boc16}. However,
these complex dynamical behaviours have so far been observed and
studied only in networks of phase oscillators.

In this article, we show how higher-order interactions
have a significant effect also on the transitions occurring
in globally coupled systems of spiking oscillators. More
specifically, we show how higher-order interactions cause
the emergence of solitary states in both directions along
the route to and from synchronization. We then characterize
the specific effects of the different orders of interaction
on the phenomenology of these systems.

\section{The model}
We study a system of globally coupled FitzHugh-Nagumo oscillators
in the presence of both pairwise and second-order interactions. Its
general evolution is described by the following system:
\begin{equation}\label{FvdP_tanh}
 \begin{aligned}
  \varepsilon\dot x_i &= x_i - \frac{x_i^3}{3} - y_i + \frac{\sigma_p}{N}\sum_{\substack{j=1\\j\neq i}}^N\tanh(x_j-x_i)\\
  &\quad + \frac{\sigma_h}{2N^2}\sum_{\substack{j=1\\j\neq i}}^{N}\sum_{\substack{k=1\\k\neq j}}^N\tanh(x_j+x_k-2x_i)\:,\\
  \dot y_i &= x_i + a_i - by_i\:.
 \end{aligned}
\end{equation}
Here, $x_i$ and $y_i$ are the state variables
of the $i$th~oscillator, the parameter $\varepsilon \ll 1$
is responsible for the time scale separation of the state variables, 
$N$ is the number of oscillators, and $\sigma_p$
and $\sigma_h$ are the strengths of pairwise
and higher-order interactions, respectively.
We consider the system in Eq.\eqref{FvdP_tanh} in the particular
case of $b=0$, so that the equations are akin to those of a van~der~Pol
model. Note that, in this interpretation, $\varepsilon$ is the inverse
of the nonlinear damping. Furthermore, we consider the oscillators
to be slightly different from each other, by imposing
that $a_i$ be random variables following a normal
distribution with mean~$0.5$ and variance~$0.001$. 

The model as defined involves nonlinear couplings
in terms of the tanh function. This specific choice
allows one to account for a saturation effect for
large values of the function argument, a behaviour
that is often desired when modelling biological processes
such as neural cell dynamics~\cite{Sni20,Hej21}.
Additionally, the use of the hyperbolic
tangent is quite convenient for a twofold reason.
First, we cannot use a linear coupling to model higher-order
interactions, as any linear function would factor
into a sum of traditional pairwise couplings:
\begin{multline}
 \sum_{\substack{j,k=1\\i\neq j\neq k}}^N\left(\alpha_i x_i+\alpha_j x_j+\alpha_k x_k\right) = \sum_{\substack{j=1\\j\neq i}}^N\left(\alpha_j x_j + \frac{\alpha_i}{2}x_i\right) +\\ \sum_{\substack{k=1\\k\neq i}}^N\left(\alpha_k x_k + \frac{\alpha_i}{2}x_i\right)\:.
\end{multline}
Moreover, close to~0, the hyperbolic tangent is equal
to the identity up to second order. Thus, when complete
synchronization is achieved and the differences between
the states of the oscillators almost vanish, the effect
of the coupling is well approximated by a linear interaction,
allowing one to apply powerful analytical methods to
investigate the phenomena observed.

\begin{figure}[t]
 \centering
 \includegraphics[width=0.45\textwidth]{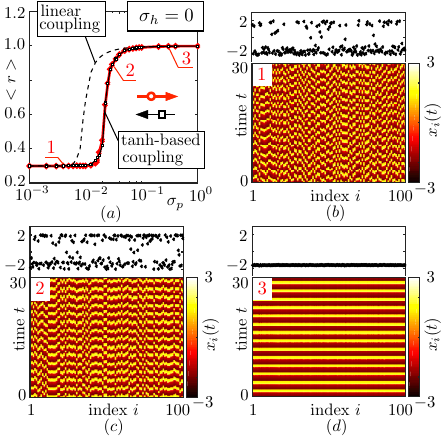}
 \caption{\textbf{The transition to synchronization is continuous in
 the absence of higher-order interactions.} (a) The time-averaged order
 parameter~$\la r\ra$ changes continuously with the two-body interaction
 strength~$\sigma_p$. The forward curve (solid red line) and the backwards
 one (solid black line) can be superimposed, and no hysteresis cycle
 occurs. The character of the transition remains the same if one uses
 a linear coupling, rather than the nonlinear one of Eq.~\eqref{FvdP_tanh}
 (dashed black line). Note the logarithmic scale for~$\sigma_p$. (b)--(d)
 Space-time snapshots of the system, taken at the points 1--3 in panel~(a),
 illustrate how, as $\sigma_p$ increases above a critical value, the
 oscillators achieve global synchronization. In these panels, point~1
 corresponds to $\sigma_p=0.004$, point~2 to $\sigma_p=0.02$ and point~3
 to $\sigma_p=0.4$. The upper insets show the final states of the system,
 which are used in the simulations as the initial conditions for the
 next points.}\label{pairwise}
\end{figure}
We study the system via numerical simulations,
using $N=100$ oscillators and $\varepsilon=0.01$,
starting from uniformly distributed random initial
conditions $x_i\in[-1.5, 1.5]$ and $y_i\in[-1, 1]$.
The steady state of a run for a given value of
the coupling strengths is used as the initial
state of the next run. At every time step, we
compute the value of the order parameter $r(t)=\frac{1}{N}\left|\sum_{i=1}^{N}\mathrm{e}^{\mathrm i\theta_i}\right|$,
where $\theta_i=\arctan(\frac{y_i}{x_i})$ is the
angle of the $i$th oscillator. The range of the
time-averaged order parameter $\la r\ra$, averaged
over the steady state of the system, is between~0
and~1, with~0 corresponding to an anynchronous
state, and~1 to fully synchronized oscillations.
Additionally, we also compute the mean phase velocity
for each oscillator, $\omega_i=2\pi\frac{M_i}{\Delta T}$,
where~$M_i$ is the number of complete rotations
around the origin performed by the $i$th oscillator
during a time interval of size~$\Delta T$.

\section{Results}
\subsection{Exclusively pairwise interactions}
\begin{figure}[t]
 \centering
 \includegraphics[width=0.45\textwidth]{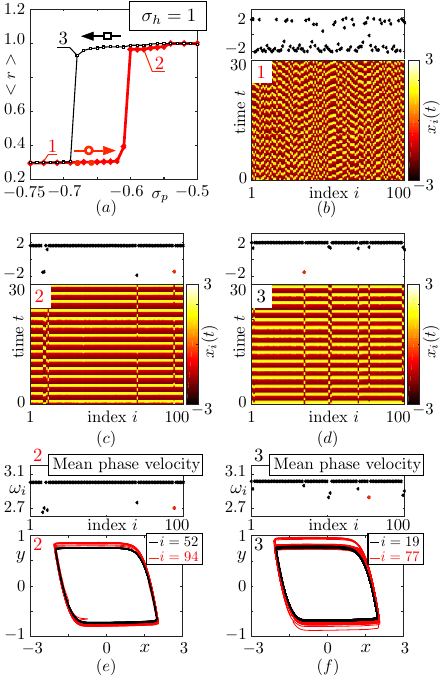}
 \caption{\textbf{Higher-order interactions make the synchronization
 transition discontinuous and induce solitary states.} (a) A hysteresis
 cycle appears in the transition to synchronization: the system switches
 to a synchronized state at $\sigma_p=-0.6$ (red line), but it reverts
 back to  the anynchronous state only when $\sigma_p$ is lowered to~$-0.69$
 (black line), indicating that the transition is first-order. (b)--(d)
 Space-time snapshots of the system, taken at the points 1--3 in panel~(a),
 show the appearance of solitary states on the route to synchronization
 and to desynchronization. In these panels, point~1 corresponds to $\sigma_p=-0.73$,
 point~2 to $\sigma_p=-0.58$ and point~3 to $\sigma_p=-0.68$. The upper
 insets show the final states of the system, which are used in the simulations
 as the initial conditions for the next points. (e)--(f) The phase portraits
 of the synchronized oscillators (black) are limit cycles, whereas those
 of the oscillators corresponding to the solitary state (red) are open
 trajectories. The upper insets show the mean phase velocities of each
 individual oscillator. Panel~(e) corresponds to point~2 and panel~(f)
 to point~3.}\label{higher}
\end{figure}
We start by considering a system with only pairwise interaction,
i.e., with $\sigma_h=0$. In such a case, a continuous transition
between synchronous and asynchronous dynamics occurs as a function
of the interaction strength. As shown in Fig.~\ref{pairwise}~(a),
no hysteresis appears in the time-averaged order parameter, and
the forward and backward curves are superimposable. The space-time
snapshots of the system, shown in Fig.~\ref{pairwise}~(b)--(d),
illustrate the transition to synchronization, with all the oscillators
evolving in-phase for $\sigma_p$ greater than a critical value.

Note that the nonlinearity of the coupling does not have a profound
impact on the transition regime. In fact, as shown in Fig.~\ref{pairwise}~(a),
the only effect caused by the specific functional form of the coupling
is merely a shift of the curve of~$\la r\ra$ as a function of~$\sigma_p$.

\subsection{Impact of second-order interactions}
\begin{figure}[t]
 \centering
 \includegraphics[width=0.45\textwidth]{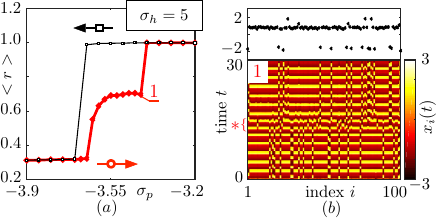}
 \caption{\textbf{Stronger higher-order interactions suppress solitary
 states during the transition to desynchronization}. (a) When $\sigma_h$
 is increased to~5, solitary states no longer appear in the backward
 transition (black line), whereas they still do in the forward one (red
 line), where they are a larger fraction than at lower values of~$\sigma_h$.
 (b) The time evolution of the system at the point~1 in panel~(a), corresponding
 to $\sigma_p=-3.425$, shows that, when solitary states appear, the specific
 oscillators that belong to them may change in time, while leaving their
 number unaltered. The moment of one such change is marked with a~$\ast$.
 The upper inset shows the final state of the system.}\label{evenhigher}
\end{figure}
When three-body interactions are introduced in the system,
a hysteresis cycle, shown in Fig.~\ref{higher}~(a), appears
in the transitions between synchronous and asynchronous states.
This indicates that the transition is no longer critical,
but rather it is a first-order one, and the possibility of
coexistence of stable incoherent and synchronized states is
expected. We note, however, that the forward and backwards
transitions are both accompanied by the occurrence of partial
synchronization with the appearance of solitary states, as
shown in Fig.~\ref{higher}~(b)--(d).

This shows that, as $\sigma_p$ increases beyond
the transition point, the system does not switch
directly to global synchronization. Rather, an
initial regime of partial synchronization is reached
such that most oscillators are in phase, while
the rest are not synchronized. Eventually, when
$\sigma_p$ increases further, the oscillators achieve
complete synchronization. The behaviour is qualitatively
unchanged, but reversed, when studying the backwards
transition. Notably, the solitary states are stable,
and they occur over a range of values of~$\sigma_p$,
showing their robustness.

The solitary states are also characterized
by outliers in the mean phase velocities,
as shown in Fig.~\ref{higher}~(e)--(f). Moreover,
the phase portraits of the synchronized oscillators
differ from those of the ones corresponding
to solitary states. Specifically, the former
are limit cycles, whereas the latter are slightly
irregular, non-closed trajectories.

\begin{figure}[t]
 \centering
 \includegraphics[width=0.45\textwidth]{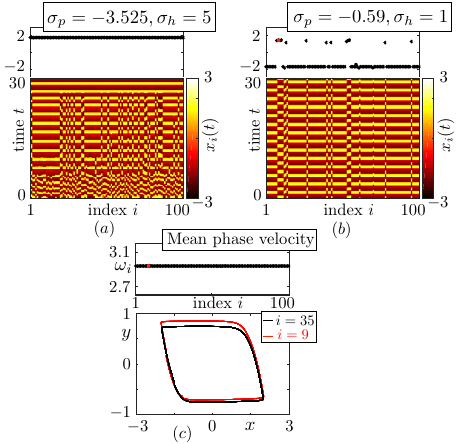}
 \caption{\textbf{Solitary states occur also for identical oscillators.}
 (a)-(b) Space-time snapshots of the system with identical oscillators at
 different values of~$\sigma_p$ and~$\sigma_h$ show the appearance of stable
 solitary states that evolve towards complete synchronization if the interaction
 strenghts are sufficiently high. The upper insets show the final states
 of the system. (c) The phase portrait of an oscillator in the synchronized
 state (black) differs from that of one in the solitary state (red), even
 though their mean phase velocity is the same (upper inset). The parameters
 have the same values as in panel~(b).}\label{identical}
\end{figure}
As the strength of higher-order interactions increases,
the forward and backward transitions exhibit a dramatic
change in behaviour. On the one hand, the desynchronization
transition becomes completely abrupt, with the suppression
of any solitary state, as shown in Fig.~\ref{evenhigher}.
On the other hand, the forward transition still features
solitary states, but with fundamentally different properties.
In particular, the specific oscillators that participate
in the solitary states changes over time. This is in stark
contrast to what happens at lower values of~$\sigma_h$,
where, instead, the oscillators that constitute the solitary
state remain unchanged. Note that the occurrence of such
a change of dynamical regime of some oscillators does not
significantly affect the fraction of asynchronous ones.
As a result, the value of the time-averaged order parameter
does not undergo large fluctuations as time progresses. 

A further remarkable feature of the transition
to synchronization shown in Fig.~\ref{evenhigher}~(a) consists
in the emergence of an intermediate  state with $\la r\ra\approx 0.7$,
which is stable for a range of coupling strengths $\sigma_p \in [-3.55, -3.425]$.
The transition to synchronization finally occurs discontinuously
at $\sigma_p \approx -3.4$. This effect is very reminiscent of
the single-step first-order transitions accompanied by heterogeneous
nucleation in adaptive dynamical networks~\cite{Fia23,Yad24}.

\begin{figure}[t]
 \centering
 \includegraphics[width=0.45\textwidth]{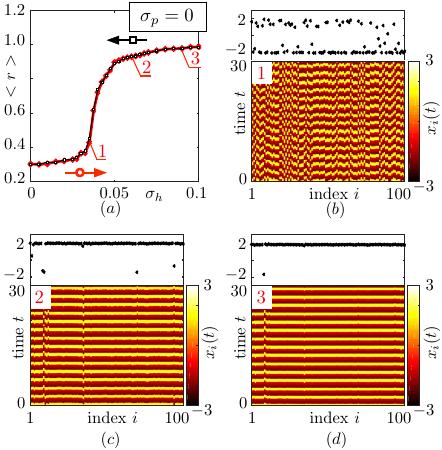}
 \caption{\textbf{Solitary states are a higher-order effect.}
 (a) The synchronization transition in the absence of pairwise
 interactions is continuous, with the forward curve (red) and
 the backwards one (black) that are superimposable. (b)--(d)
 Space-time snapshots of the system, taken at the points 1--3
 in panel~(a), illustrate the appearance of solitary states.
 In these panels, point~1 corresponds to $\sigma_h=0.0375$, point~2
 to $\sigma_h=0.06$ and point~3 to $\sigma_h=0.09$. The upper
 insets show the final states of the system, which are used in
 the simulations as the initial conditions for the next points.}\label{onlyhigher}
\end{figure}
To confirm that the solitary states are not just an artifact
due to the inhomogeneity of a system of non-identical oscillators,
we study the system when all the oscillators are forced to be
identical, which corresponds to imposing $a_i=0.5$. The snapshot
shown in Fig.~\ref{identical}~(a) shows that, at sufficiently
high~$\sigma_p$ and~$\sigma_h$, the solitary state eventually
evolves into complete synchronization. However, at lower values
of the interaction strengths, the solitary states are stable
in time, as shown in Fig.~\ref{identical}~(b). Similarly to
the case of non-identical oscillators, the phase portraits of
the attractors corresponding to the synchronized state and
those in the solitary one, depicted in Fig.~\ref{identical}~(c),
differ from each other. However, their mean phase velocities
are the same, unlike what happens when the oscillators are
not identical, and similarly to the states
first reported in Ref.~\cite{Fra22}. This phenomenology
allows us to conclude that the occurrence of solitary states 
does not depend on the existence of dynamical differences 
amongst the oscillators, but rather it is a result of the 
presence of higher-order interactions in the system.

To better understand the role of both orders of interaction,
we then consider the system in the absence of pairwise interactions,
which corresponds to $\sigma_p=0$. In this case, the results
reported in Fig.~\ref{onlyhigher}~(a) show that the synchronization
transition is continuous, with no bistability or hysteresis
cycle. Nonetheless, solitary states still do occur, as evidenced
by the the space-time snapshots in Fig.~\ref{onlyhigher}~(c)--(d),
which strongly suggest that their appearance is inherently a
higher-order effect, rather being caused by the simultaneous
presence of first-order and second-order interactions.

\begin{figure}[t]
 \centering
 \includegraphics[width=0.45\textwidth]{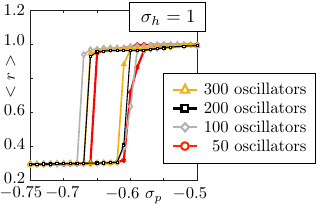}
 \caption{\textbf{The character of the synchronization transition
 is unaffected by system size.} Systems with different numbers of
 oscillators still undergo a first-order transition to synchronization,
 and the only changes are minor shifts in the transition points.}\label{sizes}
\end{figure}
To verify whether the phenomenology observed can be ascribed
to finite-size effects, we study the synchronization transition
for different system sizes. This is particularly relevant since,
in networks with higher-order interactions, this transition
has been recently reported to be strongly affected by stochastic
fluctuations for small system sizes~\cite{Sum24}. In our case,
the number of oscillators in the system only causes minor qualitative
changes in the transition, but its first-order character, shown
in Fig.~\ref{sizes}, and the presence of solitary states remain
unchanged.

\section{Conclusions}
The interplay of pairwise and second-order interactions
in systems of oscillators allows the occurrence of abrupt
transitions between the synchronized regime and the asynchronous
one, with the presence of hysteresis and bistability whereby
both coherent and incoherent states coexist. Previously,
such effects were investigated in networks of phase oscillators,
which are one of the simplest model for studying synchronization.
Here, we have shown how these transitions can occur also
in the presence of more complex dynamics, using a system
of FitzHugh-Nagumo oscillators as a case study.

Additionally, we have shown how solitary states can appear
in either transition direction under specific circumstances.
These happen in systems of identical and non-identical elements,
and they occur regardless of the strength of the pairwise
interactions. In fact, they still emerge even when two-body
interactions are entirely neglected. Conversely, in the absence
of higher-order interactions, solitary states are suppressed.
This indicates that such states are an inherent higher-order
effect in spiking oscillators, and they are not associated
with the homogeneity of the system or with its size. Note
that, unlike solitary states that were previously reported on an example 
of coupled phase oscillators ~\cite{Ska19},
which always appear only in the forward transition, those
we studied here can occur in both the transition to synchronization
and in that to desynchronization, provided that the coupling
strengths are below a critical value.

A question that arises naturally
is that of the type of the dynamics of the solitary
states. Here, we analyzed phase portraits and trajectories
obtained from different initial conditions. This
allows us to state that small changes in initial
conditions yield differences in the solitary state
trajectories that become larger over time. For this
reason, we suspect the solitary state dynamics to
be chaotic. However, a rigorous answer to the question
of the character of the solitary states would require
a deeper study involving the analysis of Lyapunov
exponents and correlation function, which we intend
to carry out in the near future.

A related question is that of the dynamical
relation between the solitary states we observe and synchronization.
In our case, the solitary states coexist with the synchronous
state and, in this sense, the observed regimes are fully consistent
with classical solitary states. Therefore, it would be possible,
in principle that their dynamics be actually transient, and
that they eventually collapse onto a fully synchronous state.
In our study, we have observed that the solitary states can
disappear and arise again involving the same oscillators. Thus,
we can conclude that they represent a robust oscillatory regime
and correspond to a particular attractor. We plan to carry out
a more detailed analysis of these aspects in future works,
since it will require the study of stochastic dynamics to check
their robustness against random perturbations, and of the differences
of evolution caused by changes in the initial conditions to identify
the basins of attraction.

Concluding, our results constitute a step forward
in the understanding of solitary states and provide
an additional piece of evidence for the existence
of inherently higher-order effects. Note
that solitary states in spiking oscillators are usually
discussed in the context of pairwise interactions,
often with nonlocal~\cite{Fra22,Ryb19} and global~\cite{Sch22}
couplings. Focusing on higher-order effects, our
research complements the existing studies, providing
observations of solitary states in systems with interactions
beyond pairwise. Relating multi-body interactions
with critical phenomena and dynamical transitions,
our work highlights the relevance of accounting for
higher-order networks when modelling complex systems.

\section*{Acknowledgements}
V.S. acknowledges support by the Russian Science Foundation (project No.~24-72-00054).
S.B. and S.J. acknowledge VAJRA grant VJR/2019/000034.
S.B. also acknowledges support from the project No.~PGR01177 of the Italian Ministry of Foreign Affairs and International Cooperation.
C.I.D.G. acknowledges funding from the Bulgarian Ministry of Education and Science, under Project No.~BG-RRP-2.004-0006-C02.

\end{document}